# Sea-level and summer season orbital insolation as drivers of Arctic sea-ice

Claude Hillaire-Marcel[1], Anne de Vernal[1], Michel Crucifix[2]

1. Geotop-UQAM, Montreal, Canada
2. UC-Louvain, Louvain-la-Neuve, Belgium

**Abstract**

The sea-ice cover of the Arctic Ocean is an important element of the climate and ocean system in the Northern Hemisphere as it impacts albedo, atmospheric pressure regimes, $CO_2$-exchange at the ocean/atmosphere interface as well as the North Atlantic freshwater budget and thermohaline circulation [1]. Due to global warming, the Arctic sea-ice cover is presently evolving at an unprecedented rate towards full melt during the summer season, driving the so-called "Arctic amplification" [2]. However, the Arctic sea-ice has also experienced large amplitude variations, from seasonal to orbital (Milankovitch) time scales, in the past. Recent studies led to suggest that whereas insolation has been a major driver of Arctic sea-ice variability through time, sea-level changes governed the development of "sea-ice factories" over shelves (Figure 1), thus fine-tuning the response of the Arctic Ocean to glacial/interglacial oscillations that is slightly out of phase compared to lower latitudes [3,4]. We discuss below how insolation and sea-level changes may have interacted and controlled the sea-ice cover of the Arctic Ocean during warm past intervals and how they could still interfere in the future.

## The "Milankovitch" pacing of the Arctic Ocean climate and sea-ice

The insolation of the Summer solstice at 65°N is commonly used as a timer for ice growth and decay in the Northern Hemisphere [5]. However, because the Arctic sea-ice records its minimum extent in September [6], the heat accumulated and transferred from lower latitudes over the whole summer season may be more relevant. Through time, the pacing of summer vs solstice precessional maxima are close (Figure 2a), but their relative amplitude differs significantly from one peak to another. Moreover, the two curves depict a timing offset of a few thousand of years (Figure 2b). Prior to looking deeper into these features, the choice of the 65°N latitude (~ Iceland) for estimating a "Summer" insolation forcing of the Arctic Ocean sea-ice cover deserves examination. On one hand, the pattern is not significantly different with Summer insolation at 80°N. It would be slightly lower (-22.7±2.7 $W.m^{-2}$; ± 1$\sigma$) but nearly parallel. On the other hand, an important vector of heat towards the Arctic Ocean is the North Atlantic Water (NAW) mass (Figure 1), which flows northwards from the Nordic seas through the Fram Strait and the Barents Sea, then circulates below the low-salinity surface layer of the Arctic Ocean. The 9 ± 2 Sv of NAW entering the Arctic Ocean provide a heat flux of about 40 TW [7], which plays a determinant role in the melting of Artic sea-ice in summer [8]. This heat flux is practically set at the latitude of the northern North Atlantic Ocean, mostly in relation with the subpolar gyre [9] centered at about 60°N. The choice of 65°N seems thus relevant and permits a direct comparison with the commonly used insolation curve based on the Summer solstice at this very latitude.

The 2.5 ka time-offset observed between the Summer and June 21$^{st}$ insolation curves (Figure 2b), is a most interesting feature. It is probably not critical during insolation minima, due to the extensive ice sheets, but becomes critical between maxima such as that of the present Holocene interglacial. As recorded in marine sedimentary cores from the Central Arctic Ocean, the evidence for significant sediment flux and productivity during the Holocene is dated at ca 8-9 ka [10], which is practically in phase with the mean Summer season insolation peak but lags by ~ 2.5 ka the Summer solstice insolation peak (Fig. 2b). The dated foraminifera from this interval were mostly produced in the seasonally open sea-ice factories of the circum-Arctic shelves, then largely submerged as a result of the postglacial sea-level rise. These populations, almost exclusively composed of the planktonic *Neogloboquadrina pachyderma,* related to heterotrophic production in the subsurface-mesopelagic layer [11]. They developed during the productive sea ice-free season, were carried with currents and/or uploaded into sea-ice with suspended sediment, then transported and deposited along the major sea-ice transit routes (TransPolar Drift and Beaufort gyre of Fig. 1), notably along the Lomonosov Ridge [4]. The high Summer insolation of the early Holocene combined with the high nutriment availability linked to costal erosion due to the rising sea level (SL), triggered high primary productivity [12].

Summer insolation including the duration, height and succession of the maxima caused by climatic precession need to be taken into consideration (Fig. 2a). A comparison of recent interglacials leads to infer specific responses of the Arctic Ocean. For example, the long marine isotope stage (MIS) 11 has led to a significant retreat of the Greenland Ice Sheet [13] and to sea-level rise about 20 m above that of the present interglacial. However, its Summer insolation was much below that of all other interglacials of the last million years. Warm conditions in the Arctic Ocean during MIS 11 would thus be unlikely, which is consistent with paleoceanographic data from the northern North Atlantic [14]. Model experiments forced with insolation led to similar inferences about MIS 11 [15]. Unfortunately, the only attempt at the documenting of MIS 11 conditions in the Arctic Ocean

[16] from marine records cannot be used due to uncertainties in their chronostratigraphy. Based on independent chronostratigraphic evidence [10, 1], the interval assigned to MIS 11 in [16] can be re-assigned to an older "warm" interval, probably MIS 15 according to the position of the Brunhes/Matuyama magnetic reversal in the sedimentary sequence [16, 17].

Another feature of the Summer insolation deserves attention. It concerns the relative amplitude of the precessional peaks of MIS 7 vs MIS 5. The insolation maxima of MIS 7a and 7e suggest warmer and more open Arctic Ocean than those of MIS 5a and 5e. This is supported by central Arctic marine records which document a longer duration and more effective sea ice rafting deposition during MIS 7 than MIS 5e [10,1]. Back in time, the MIS 15 evoked above presents much analogy with MIS 7.

From an Arctic Ocean perspective, the Milankovich forcing of the whole Summer mean insolation seems a robust driver although it is not fully carried in the literature, especially with respect to paleoceanographic conditions during MIS 5e and 11, both weakly recorded in central Arctic Ocean by seasonal sea-ice proxies.

**The sea-level control of the Arctic sea-ice dynamics**

Sea-ice mostly forms over the large Arctic Ocean shelves, especially in the Laptev and East Siberian seas (Fig. 1a). During glacials, when sea-level was about 120-130 m below its present level [18], Bering Strait, the shallow gateway (~ 50 m deep) linking the Arctic and the Pacific oceans was emerged and closed, while the Barents Sea and the Canadian Arctic Archipelago gateways were practically closed by ice. The Arctic Ocean was then a deep semi-enclosed basin, with a narrow connection to the North Atlantic through the Fram Strait [19]. During glacial stages, the Arctic Ocean was covered by a thick ice shelf with ice streaming from surrounding glaciers (Fig. 1b) [20]. Following deglaciation, ice melting and the rising sea-level led to the re-opening of all gateways, with a progressive submergence of Arctic shelves (Fig. 2a), thus fostering production of sea-ice. "Warm" NAW then entered the Arctic Ocean through both the Fram Strait and the Barents Sea, with impact on the seasonal melting of sea-ice. On the Pacific facade, the opening of Bering Strait permitted low-salinity waters from the North Pacific to enter the Arctic Ocean, impacting its upper water mass stratification and sea-ice production over shelves. Today, the Pacific flux weights up to 40% of the freshwater budget of the Arctic Ocean [21], aside bringing about 20 kmole.s$^{-1}$ of silicium into the Western Arctic [22], hence also impacting diatom productivity, thus heterotrophic plankton production.

At the onset of the present interglacial, the Barents Sea gateway opened at ~12 ka BP [23]. The Bering Strait followed at ~ 11 ka BP [24]) when the global SL was about still ~ 50 m below the modern one [18]. However, maximum flow conditions through Bering Strait were only achieved in mid-Holocene time [25], when the mean global sea-level reached its Holocene maximum [e.g., 24]. Data from the Chukchi Sea and the Western Arctic indicate that near-similar modern-like sea-level and flow conditions through Bering Strait were only reached some 4 ka ago [25], i.e., at the onset of the Arctic cooling trend towards the neoglacial [4]. The "downstream" export of Arctic Ocean low-salinity water, in part originating from the Pacific, then impacts the AMOC [26]. As a consequence of sea-level rise, not only the Arctic Ocean shows a specific interglacial timing and response, but its final "interglacial" feedback on southern latitude, delayed by several thousand years, could perhaps explain apparently "out-of-phase' instabilities .

Sea-level seems thus a joker here, exerting a strong control over the Arctic Ocean hydrography and subsequently the North Atlantic thermohaline circulation. The difficulty now is to document how sea-level may have controlled, combined with the Summer mean insolation, the conditions of past "warm" intervals in the Arctic Ocean. Usually, the stacked $\delta^{18}O$ of benthic foraminifers from deep-sea cores is used as a proxy for sea-level changes through time [27]. However, taking into account the long delay, up to ~ 5 ka if not more, for a full transfer of the glacial meltwater isotope signal to the deep ocean [28], adding the smoothing effect of stacking and the complex $^{18}O$-salinity-ocean-ice volume relationships [29], we can assume that the Milankovitch-tuned benthic $^{18}O$ stack cannot provide a time frame accurate enough to determine effective time-offsets for the achievement, in the Arctic Ocean, of "full interglacial" conditions that include minimum land ice volume, maximum sea-level and full isostatic equilibrium. In addition, assuming that maximum rates of sea-level changes would be reached within 2 kyr following the continental ice decay [30], sea-level rises linked to fast but short melting events are unlikely to be captured by the oceanic $^{18}O$-proxy of sea-level.

**The Arctic Ocean during past interglacials**

Whatever the limitations of the marine $\delta^{18}O$-stack as a proxy for sea level (SL) changes, it has to be used for a reasonable assessment of the status of the Arctic sea-ice combined with the Summer insolation parameter for earlier interglacials. As illustrated Figure 2a, we define thresholds for the two parameters, based on the summer mean insolation value at the onset of the present interglacial (the Holocene), and the $\delta^{18}O$-value corresponding to a SL ~ 50 meters below its present elevation, when shelves were already largely-submerged and at the time Bering Strait opened. During MIS 5, only the optimum (MIS 5e) shows conditions suitable for "open Arctic Ocean" (with probable seasonal sea-ice melting), thus a short, ≤ 10 kyr interval, that is unlikely to have left more evidence in sedimentary records than the present Holocene with about 1 or 2 cm-thick layers [3,4]. In contradiction, during the ~ 60 kyr long MIS 7, several intervals with open conditions were likely. Noteworthy is the fact that the insolation peak of MIS 7e was significantly higher than that of MIS 5e, practically equivalent to that of MIS 7a. Accordingly, a much thicker MIS 7 layer can be expected in sedimentary records from the central Arctic Ocean. The strongest evidence for intense sea-ice rafting during MIS 7 is provided by its high residual content in excess $^{230}$Th ($^{230}$Th$_{xs}$), the $^{230}$Th produced by the dissolved uranium in the ocean water [31]. Its scavenging in the water column indicates active sea-ice and brine production, thus seasonally open conditions [3],.

Another striking feature is the unlikeness of any strong sedimentary recording of MIS 11, as its mean Summer-season insolation remained very low in comparison with those of MIS 7 and MIS 15, notably (Figure 2a). MIS 11 distinctive features have been discussed in many papers, either based on data [14] or modelling [13]. From an Arctic Ocean perspective, the closest "warm analogue" of the Holocene would be MIS 9e rather than MIS 11.

On the contrary, evidence for a short "open sea-ice" interval during MIS 3 is provided by high $^{230}$Th$_{xs}$ values in sedimentary cores [3]. The event was likely too short to be recorded by the marine $\delta^{18}$O-stack. However, several studies indicate that mean SL could have been close to 40 below its present elevation some 55 ka BP [32], whereas the mean Summer-season insolation at that time culminated well above that of MIS 11. This observation suggests that a high mean Summer-season insolation constitutes a necessary condition for an open Arctic Ocean since MIS 11 does not seem well recorded despite high SL and duration [27]. Two arguments support the existence a short open Arctic interval by ~ 55 ka BP. The first one deals with model experiments suggesting a deep (≥ 2000 m) and high (~ 20 Sv) convection at ~ 60°N in the North Atlantic [33], thus conditions not unsimilar present ones. The second argument is provided by an isotopic and palynological study of interbedded ice wedges and peat layers near the Laptev Sea coast, which suggests milder conditions during MIS 3 than during MIS 5 [34]. Hence, a short MIS 3 "warm" Arctic Ocean event seems plausible. Older intervals with similar insolation and high sea-level values deserve some attention (e.g., MIS 9a and 13a; Figure 2a).

# The Arctic Ocean: from triggers to feedbacks, from past to future

Feedbacks of the specific climatic response of the Arctic Ocean towards climate/ocean conditions at lower latitudes of the North Atlantic are several but the most important are i) the effect of sea-ice on albedo and latitudinal pressure gradients and ii) the impact of freshwater export on the AMOC. In the Arctic Ocean, slightly out of phase and shorter "warm" intervals with an early productivity peak, then a later nearly stabilized freshwater budget, may have impacted "interglacial" conditions at lower latitudes. Presently, the amplified warming of the Arctic under the present greenhouse-forcing is irreversible [35]. A close (pseudo) analogue of its near future could be found in the late Pliocene, when the atmospheric $pCO_2$ was about 400 ppmV [36] and the boreal forest reached Ellesmere Island (Fig. 1b) on the Canadian margin of the Arctic Ocean [37]. Then, mean annual temperature on Ellesmere Island were probably close to those observed today in the Hudson Bay, about 2500 km southward. If we push the analogy further, the ~ 200 days.year$^{-1}$ of sea-ice presence in the modern Hudson Bay could well represent the fate of the Arctic ocean in the near future: a seasonal-only sea-ice cover with higher productivity [38]. On a longer term, the ongoing melting of the Greenland and Western Antarctica ice sheet will lead to a sea-level rise of several meters, deepening the Bering Strait, thus lowering further the salinity budget of the Arctic Ocean, aside a predicted runoff of circum-Arctic river [39]. Lower North Atlantic-Arctic pressure gradients, a sluggish AMOC are the most immediate expected responses. Given the rapid ongoing changes in the cryosphere, it is thus also highly relevant to take into consideration the impact of sea-level rise on the Arctic freshwater budget, sea-ice dynamics and feedbacks in the ocean-climate system, for climate modeling [40] and a proper assessment of the Earth's future climate.

**Figure captions**

Figure 1 - A - Outline of the modern "interglacial" Arctic Ocean (base map: IBCAO/NCEI); White arrows: major sea-ice drift routes; blue arrows: low salinity water currents; blue/white grading arrows: Arctic "freshwater" export (low salinity currents and ice export); dark to light brown arrows: main sub-surface North Atlantic Water (NAW) route; AMOC: Atlantic Meridional Overturning Circulation. - B - Sketch of the Arctic Ocean with reduced shelves (thus sea-ice factories) under full glacial conditions with exchanges restricted to the Fram Strait gateway (white arrow: "freshwater" export brown arrow: NAW influence).

Figure 2 - A - Comparison of the Summer solstice and mean Summer-season insolation at 65°N with the $\delta^{18}O$-stack used as proxy of sea-level changes during the last 600 ka [27]; The ~ 390 W.m$^{-2}$ brown dashed line illustrates insolation similar to that of the early Holocene, with high productivity potential ; The ~ 3.5‰ blue dashed line illustrates sea-level of the early Holocene, when Bering Strait opened; the turquoise rectangles highlight past interglacials with conditions favorable for long intervals with seasonally opened sea ice and intense sea-ice drifting over the central Arctic Ocean. MIS 7 (and likely MIS 15) show more "open" Arctic Ocean conditions in comparison to MIS 11 in particular; blue arrows indicate interstadial intervals with possibility for short "open" conditions - B - Blow up of the insolation curves for the last 100,000 years illustrating the time offset between the two insolation parameters; the mean value of the astronomical season seems to fit better with high productivity conditions in the Arctic Ocean than that of the Summer solstice.

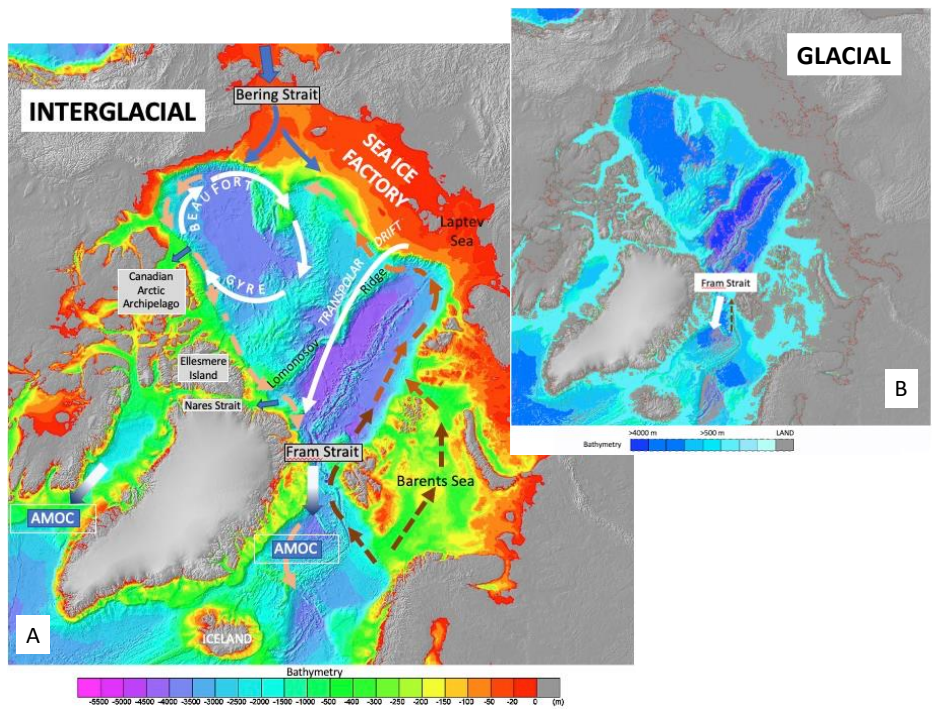

Figure 1

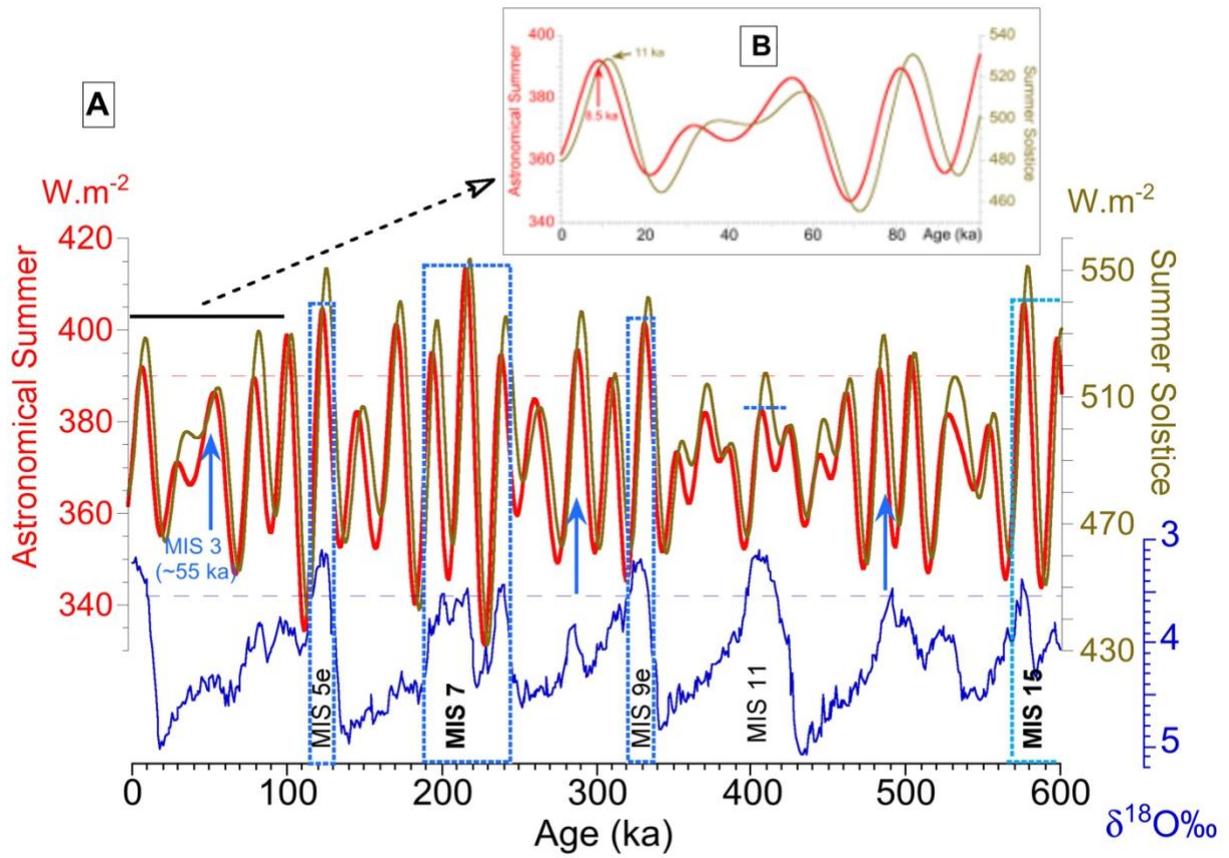

Figure 2